\documentclass[aps,prb,twocolumn,groupedaddress]{revtex4}

\usepackage{graphicx}

\begin{document}


\title{Magnetotransport in inhomogeneous magnetic fields}

\author{Tohru Kawarabayashi}
\affiliation{Department of Physics, Toho University,
Miyama 2-2-1, Funabashi 274-8510, Japan}

\author{Tomi Ohtsuki}
\affiliation{Department of Physics, Sophia University,
Kioi-cho 7-1, Chiyoda-ku, Tokyo 102-8554, Japan}

\date{\today}

\begin{abstract}
Quantum transport in inhomogeneous magnetic fields is investigated
numerically in two-dimensional systems using the equation of motion method. 
In particular, the diffusion 
of electrons in random magnetic fields in the presence of additional weak 
uniform magnetic fields is examined.
It is found that the conductivity is strongly suppressed by the 
additional uniform magnetic 
field and saturates when the uniform magnetic field becomes on the 
order of the fluctuation of the random magnetic field. 
The value of the conductivity at this saturation is found to be 
insensitive to the magnitude of the fluctuation of the random field.
The effect of random potential on 
the magnetoconductance is also discussed.
\end{abstract}


\maketitle

\section{Introduction}
Quantum transport in two-dimensional (2D) disordered systems 
in inhomogeneous magnetic fields has been 
studied extensively, in connection with the fractional 
quantum Hall effect (FQHE).\cite{HLR}
In particular, the Anderson localization\cite{LR} in 
a 2D system in random magnetic fields(RMF) with zero mean, 
which arises in the mean field theory of the fractional 
quantum Hall effect at filling factor $\nu=1/2$, has 
been studied by many authors.\cite{PZ,KZ,SN,AHK,YG,Verges,SYN,Yakubo,KO} 
From the theoretical point of view, whether such a system 
has a metallic phase or not has been an important issue.  
Systems in random magnetic fields belong 
to the 2D unitary universality class which 
is, in general, expected to have no metallic phase. If the 
system has a metallic phase, it throws a question on the 
validity of the conventional classification of the universality classes
of the Anderson transition. 
Although many numerical as well as analytical attempts  has been made to 
clarify this point, the conclusions are so far still controversial. 
On the other hand, the singular behavior of the conductance 
fluctuations\cite{OSO} and the density of states at the band center $(E=0)$ 
has been investigated and argued that these singularities are governed by
the chiral unitary universality class.\cite{Furusaki,MBF}

Apart from such a theoretical interest, 
the transport properties of 2D electron systems
in random magnetic fields are important to understand 
the experiments in the quantum Hall systems. 
In fact, the structure observed in the magnetoresistance 
at the filling factor $1/2$ has been analyzed using 
models with random magnetic fields.\cite{EMPW}
On the other hand, recently, 
two-dimensional electron systems 
in random magnetic fields have been realized experimentally 
by attaching small magnets on the layer parallel to 
the 2D electron gas in a semiconductor heterostructure.\cite{Ando}
In such a system, one can also control the strength of the 
random magnetic fields.
The magnetotransport has been measured and interestingly, it is found 
that the magnetoresistance 
in systems having random magnetic fields exhibits similar structures to that 
in the fractional quantum Hall system at the filling factor $\nu=1/2$.
\cite{Ando} 
Since the origins of the magnetic field fluctuations in these 
two systems are quite different,  
it is important to clarify the origin of this similarity 
in the magnetoresistance between these two systems.

One of the significant features of the magnetoresistance observed in 
the recent experiment\cite{Ando} 
is the increase of the magnetoresistance in two different scales of the 
magnetic field.
One is the dip structure around zero field whose width is on
the order of the fluctuation of the random field, and the other is
the positive 
magnetoresistance in a larger scale of the magnetic field
followed by the Shubnikov-de Haas oscillations.  
In recent years, the semi-classical theory for the magnetotransport 
in a smoothly varying random magnetic field has been developed,\cite{MWEPW}
where the snake state near zero magnetic field lines plays an important 
role.
In the case of a weak disorder, 
a pronounced positive magnetoresistance coming from  
a classical origin has been reported for small magnetic fields.\cite{MWEPW}
Although the semi-classical approach provides qualitatively consistent 
results with the experiments, 
it is also important to examine the transport 
property by the quantum mechanical calculations for its further 
understanding.

In the present paper, we study the magnetoconductance in the 
random magnetic field in two dimensions.
We consider a tight-binding model in random fluxes having no
spatial correlation, which has been analyzed by many authors 
for the study of the Anderson localization in random magnetic fields.
In two dimensions, it has been shown that such a model has
the large localization length 
near the band center and the diffusive behavior 
is observed in that energy regime for numerically accessible length 
scales.\cite{KO} 
Since there is no spatial correlation 
in random magnetic fields, the random magnetic field is not 
smooth and hence the validity of 
the semi-classical concepts is not obvious for the present model.
In our model, we find that the conductivity 
is strongly suppressed by adding a uniform field and also 
that it shows a saturation when the uniform field becomes on the order
of the fluctuation of the random magnetic field. 

We adopt the equation of motion method in order to 
examine numerically the diffusion of an electron in 
inhomogeneous magnetic fields. 
Adopting this method, we have an advantage that very large systems
compared with other numerical methods can be considered. 
However, within the present method,
we are able to estimate the longitudinal conductivity $(\sigma_{xx})$
only, and not the Hall conductivity $(\sigma_{xy})$. Due to this, 
we are not able to discuss the magnetoresistance directly.
As mentioned above, the 2D random magnetic field system has been shown 
to have very large localization length near the band center. 
It is then natural to assume that, in the present model, this 
regime is responsible for the metallic 
behavior observed in experiments. We therefore 
confine ourselves to the case that the Fermi energy lies near 
the band center. 
First, we discuss the transport in the random magnetic fields with zero mean 
and next, we  
examine the effect of additional uniform magnetic fields. 
The strength of the additional 
uniform magnetic field considered in the present study
is assumed to be on the same order as or 
smaller than that of the random magnetic field.   
Implications from our numerical results 
for the magnetoconductance are discussed 
in comparison with the recent experimental results.

\section{Model and Method}
In order to describe the two-dimensional system in random
magnetic fields, we consider the following Hamiltonian 
\begin{equation}
 H = \sum_{<i,j>} V \exp({\rm i} \theta_{i,j}) C_i^{\dagger}
 C_j  + \sum_{i} \varepsilon_i C_i^{\dagger} C_i
\end{equation}
on the square lattice.
Here $C_i^{\dagger}(C_i)$ denotes the creation(annihilation) operator 
of an electron on the site $i$ and $\{ \varepsilon_i \}$ denote the 
random potential distributed independently in the 
range $[ -W/2, W/2]$.  
The phases $\{ \theta_{i,j} \}$ are related to the magnetic fluxes 
$\{ \phi_i \}$ through the plaquette 
$(i,i+\hat{x},i+\hat{x}+\hat{y},i+\hat{y})$ as 
\begin{equation}
 \theta_{i,i+\hat{x}} + \theta_{i+\hat{x},i+\hat{x}+\hat{y}}+
 \theta_{i+\hat{x}+\hat{y},i+\hat{y}} + \theta_{i+\hat{y},i}
 = -2\pi \phi_i / \phi_0
\end{equation}
where $\phi_0 = h/|e|$ stands for the unit flux.
The fluxes $\{ \phi_i \}$ are also assumed to be distributed independently 
in each plaquette. The probability distribution $P(\phi)$ of the flux $\phi$ 
is given by
\begin{equation}
 P( \phi ) = \left\{ \begin{array}{ll}
                     1/h_{\rm rf} & {\rm for }\quad  |\phi /\phi_0| 
                     \leq h_{\rm rf}/2 \\
                      0 & {\rm otherwise}
                     \end{array}  \right.  . \label{dist1}
\end{equation}
The variance of the distribution is accordingly given by
\begin{equation}
 \langle \phi_i \phi_j \rangle = \frac{h_{\rm rf}^2}{12}\phi_0^2 
 \delta_{i,j} . 
\end{equation}

In order to solve the time-dependent Schr\"{o}dinger equation
numerically, we employ the decomposition formula for 
exponential operators.\cite{Suzuki} 
The basic formula used in the present paper 
is the forth order formula
\begin{equation}
 \exp ( x[A_1 + \cdots + A_n]) = S(xp)^2 S(x(1-4p))S(xp)^2 + O(x^5),
\end{equation}
where 
\begin{equation}
 S(x) = {\rm e}^{xA_1/2}\cdots {\rm e}^{xA_{n-1}/2}{\rm e}^{xA_n}
 {\rm e}^{xA_{n-1}/2}\cdots {\rm e}^{xA_1/2}.
\end{equation}
The parameter $p$ is given by $p=(4-4^{1/3})^{-1}$ and $A_1, \ldots , A_n$
are arbitrary operators.
We divide the Hamiltonian into five parts as in the 
previous paper\cite{KO} so that each part is represented 
as the direct product of $2\times 2$ matrices.
By applying this formula to the time evolution operator $U(t)\equiv 
\exp(-{\rm i}
H t/\hbar)$, we obtain 
\begin{eqnarray}
 U(\delta t) &=& U_2(-{\rm i} p \delta t/\hbar)^2 
 U_2(-{\rm i}(1-4p)\delta t/\hbar) 
 U_2(-{\rm i} p \delta t/\hbar)^2 \\
 & &+O(\delta t^5)
\end{eqnarray} 
with
\begin{equation}
 U_2(x) = {\rm e}^{xH_1/2}\cdots {\rm e}^{xH_4/2}{\rm e}^{xH_5}
 {\rm e}^{xH_4/2}\cdots {\rm e}^{xH_1/2} ,
\end{equation}
where $H=H_1 + \cdots + H_5$. It is to be noted that the $U_2$ can 
be expressed in an  analytical form while the original evolution operator 
$U$ can not be evaluated exactly without performing the 
exact diagonalization of the whole system. 
We are thus able to 
consider larger system-sizes than other numerical methods, such as 
exact diagonalization and the recursion method based on 
the Landauer formula, which is one of the advantages of the 
present method.
This method has already 
been successfully applied to the case of  
$W=0$ and $h_{\rm rf}=1$\cite{KO} as well as
to the 2D symplectic class.\cite{KO2} 

The system we consider is the square lattice of the size $999 \times 999$
with the fixed boundary condition. All the length scales are measured in
units of the lattice constant $a$.
To prepare the initial wave packet with energy $E$, we numerically 
diagonalize the subsystem ($21 \times 21$) 
located at the center of the whole system and 
take  the eigenstate whose eigenvalue is the closest to $E$ as the initial 
wave function. 
In the following, we set the energy $E$ to be $E/V = -0.5$ which is 
close to the band center.
The single time step $\delta t$ 
is set to be $\delta t = 0.02 \hbar/V$.
With this time step, the fluctuations of the expectation value of 
the Hamiltonian is safely neglected throughout the present 
simulation $(t \leq 200 \hbar/V)$.
We observe the second moment defined by 
\begin{equation}
 \langle \mathbf{r}^2 (t)\rangle_c \equiv \langle \mathbf{r}^2 (t)\rangle -
 \langle \mathbf{r} (t)\rangle^2
\end{equation}
with 
\begin{equation}
 \langle \mathbf{r}^n(t) \rangle = 
 \sum_{\mathbf{r}} \mathbf{r}^n  |\psi(\mathbf{r},t)|^2, \qquad (n=1,2)
\end{equation}
where $\psi(\mathbf{r},t)$ denotes the wave function at time $t$.
In the diffusive regime, the second moment is expected to 
grow in proportion to $t$
\begin{equation}
 \langle r^2 \rangle_c = 2dDt ,
\end{equation}
where the diffusion coefficient is denoted by $D$ and $d$ is
the dimensionality of the system. The 
diffusion coefficient $D$ is related to the conductivity 
by the Einstein relation $\sigma = e^2 D \rho$.\cite{KTH} Here 
$\rho$ denotes the density of states. In Fig. 1, the 
second moment is shown as a function of time for the case of 
$W=0$. For each value of $h_{\rm rf}$, 
five realizations of random magnetic fields are considered.
It is clearly seen that for $t \geq 50 \hbar/V$
the second moment increases in proportion to $t$, which means that 
the system is in the diffusive regime. For $t \leq 50 \hbar/V$, we  see 
the ballistic behavior, where the second moment grows as 
$\propto t^2$. By examining the behavior of the second moment,
we can clearly distinguish whether the system is in the diffusive regime
or in the ballistic regime. 
To estimate the diffusion coefficient we discard the 
data in the ballistic regime. We also do not consider the 
time scale where the wave packet reaches to the edge of 
the whole system.
Within the time scale considered in the present paper $t\leq 200 \hbar/V$,
we do not observe any sign of the saturation of the second moment 
due to the finiteness of the system.
To obtain the conductivity, we need the 
density of states too. We estimate it by the Green function 
method\cite{SKM} for strips the width of which is $12 \leq M \leq 30$.  

\begin{figure}
\includegraphics[scale=0.6]{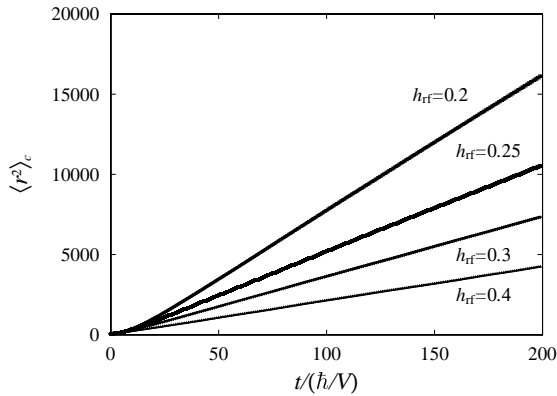}%
\caption{The second moment as a function of time 
for $h_{\rm rf}=0.2, 0.25, 
0.3$, and $0.4$. Five realizations of random magnetic fields are
considered for each $h_{\rm rf}$. \label{fig1}}
\end{figure}

\section{Numerical Result}

Let us first discuss how the diffusion 
of an electron in random magnetic fields depends on 
the strength of the fluctuation of random magnetic fields.
We estimate the diffusion coefficients from the behavior
of the second moment shown in Fig. 1. It is clearly seen that 
the diffusion coefficient becomes smaller as the fluctuation
of the random magnetic fields increases.
The density of states at $E=-0.5V$ 
is estimated to be $\rho \approx 0.178$, $0.181$,
$0.179$, and $0.179 (1/a^2V)$ for $h_{\rm rf}=0.2$, $0.25$, 
$0.3$ and $0.4$, respectively.
It depends on the strength of the
RMF very weakly near the band center.
On the other hand, the diffusion coefficient $D$ depends 
strongly  on the strength of the RMF which is 
estimated to be $4D=85.22\pm 0.03$, $54.44\pm 0.04$, $37.40\pm0.03$ 
and $21.30\pm 0.02 (a^2V/\hbar)$ for $h_{\rm rf}=0.2$, $0.25$, 
$0.3$ and $0.4$, respectively.
In Fig. 2, the conductivity, evaluated from these diffusion coefficients
and the density of states, is plotted for small values of $h_{\rm rf}$. 
Here we clearly see that 
the conductivity is inversely proportional to the square of $h_{\rm rf}$,
namely, $\sigma \propto 1/h_{\rm rf}^2$. This fact indicates that
the present regime would be well described by the 
Born approximation.\cite{EMPW} 
In the recent experiment,\cite{Ando} it is observed that the 
relative change of the resistivity by the random magnetic fields 
is in proportion to the square of the fluctuation 
of the random magnetic fields, which is qualitatively 
consistent with the present results (Fig. 2).

\begin{figure}
\includegraphics[scale=0.65]{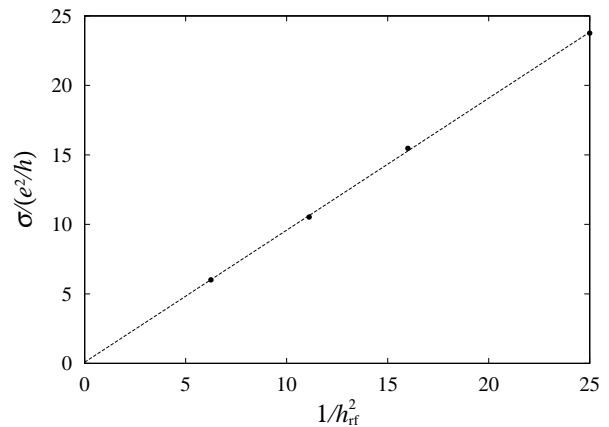}%
\caption{Conductivity in units of $e^2/h$ for $h_{\rm rf}=0.2$, $0.25$,
$0.3$ and $0.4$. \label{fig2}}
\end{figure}

Next, we apply an additional uniform magnetic field and 
examine its effect on the electron diffusion. The uniform 
magnetic field through
each plaquette of the square lattice is denoted by $\phi_{\rm uni}$.  
The total magnetic field per plaquette is then $\phi_{\rm uni}+\phi$, 
where $\phi_{\rm uni}$ is common to all the plaquette and $\phi$ is 
distributed independently as (\ref{dist1}).
The second moments for $h_{\rm rf}=0.2$ under various values of 
the uniform magnetic fields are shown in Fig. 3.
Here it is clear that the diffusion of electrons is 
strongly suppressed by the additional uniform magnetic field.
Note that the strength of the uniform field considered is smaller than the 
fluctuation of the random field $\sqrt{\langle \phi^2/\phi_0^2 
\rangle} = h_{\rm rf}/\sqrt{12}= 0.0577\ldots $. 
The diffusion coefficients and the density of states 
estimated for various values of the uniform fields 
are summarized in Table 1.
Here we see again that the density of states is not 
sensitive to the uniform magnetic field. 
We also perform numerical calculations for $h_{\rm rf}=0.3$ and
$0.4$. In Fig. 4,
the estimated conductivity is plotted as a function of the 
uniform field scaled by $\sqrt{\langle \phi^2 
\rangle}$.
For these values of 
the random magnetic field, it is 
commonly observed that the negative magnetoconductance 
occurs when the uniform field is weaker than the 
fluctuation of the random field.
It is also observed that at $\phi_{\rm uni} \approx 
\sqrt{\langle \phi^2 \rangle}$, the conductivity takes a value 
on the order of $e^2/h$ and is, interestingly, insensitive to the 
magnitude of the random magnetic field.

\begin{table}
\caption{The diffusion coefficients $D$ and the density of states 
$\rho$ for 
$h_{\rm rf}=0.2$ \label{}}
\begin{ruledtabular}
\begin{tabular}{ccc}
 $\phi_{\rm uni}/\phi_0$ & $4D$  & $\rho$\\
    0    & 85.22 $\pm$ 0.03 & 0.178 \\
   0.002 & 83.25 $\pm$ 0.03 & 0.179\\
   0.005 & 76.37 $\pm$ 0.04 & 0.179 \\
   0.01  & 59.54 $\pm$ 0.02 & 0.178\\
   0.02  & 34.61 $\pm$ 0.04 & 0.179\\
   0.04  & 14.65 $\pm$ 0.02 & 0.183 \\
   0.06  & 8.69 $\pm$ 0.02 &  0.189\\
   0.08  & 6.69 $\pm$ 0.02 &  0.177\\
   0.1   & 5.31 $\pm$ 0.02 &  0.168\\
\end{tabular}
\end{ruledtabular}
\end{table}

\begin{figure}
\includegraphics[scale=0.65]{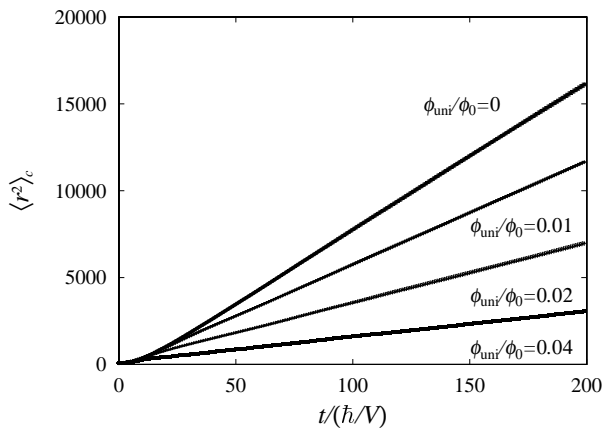}%
\caption{Electron diffusion for
$\phi_{\rm uni}/\phi_0=0$, 
$0.01$, $0.02$, and $0.04$, where $\phi_{\rm uni}$ stands for 
the strength of the uniform flux per plaquette. \label{fig3}}
\end{figure}

\begin{figure}
\includegraphics[scale=0.65]{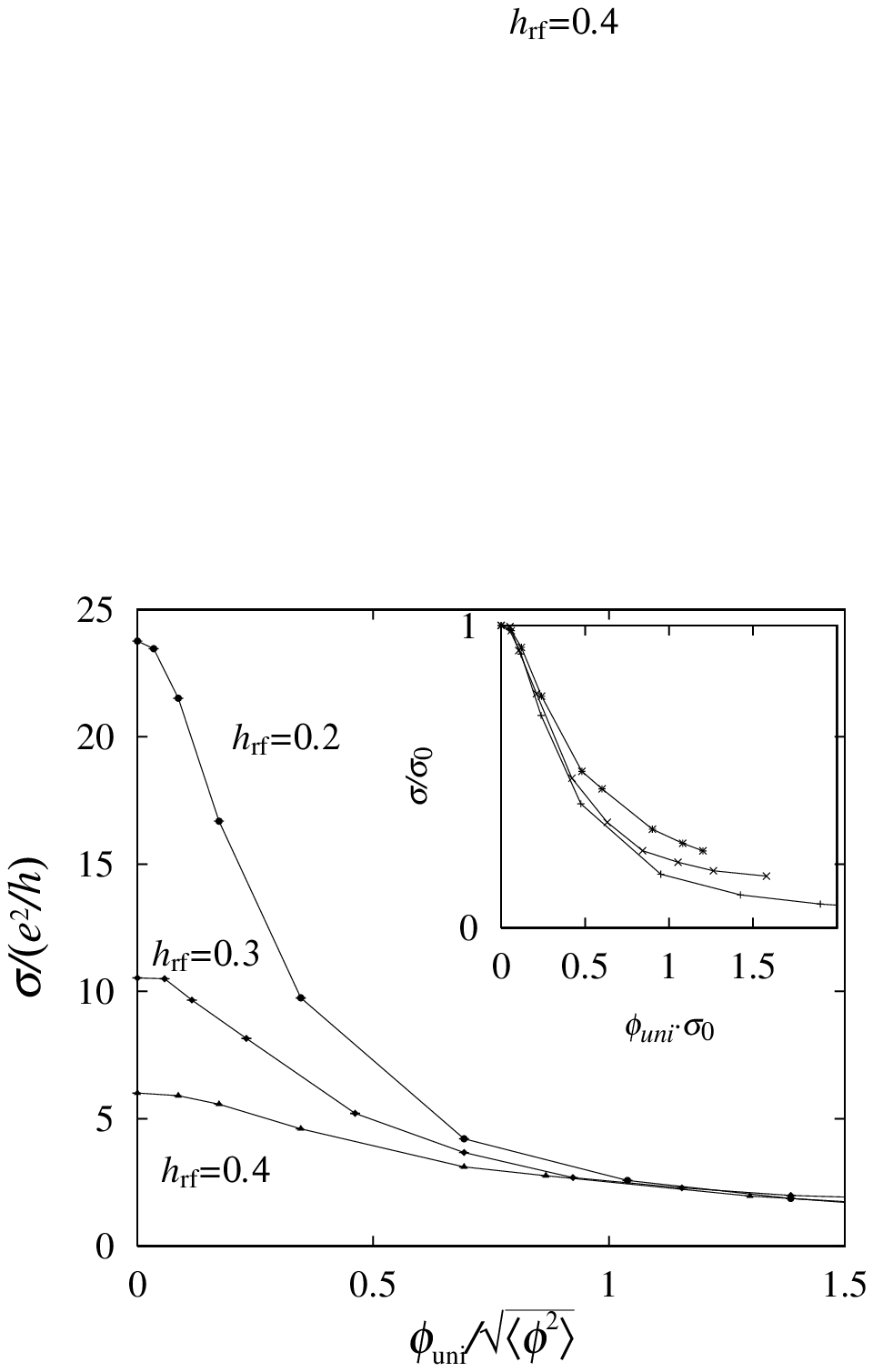}%
\caption{Magnetoconductance for
$h_{\rm rf}=0.2$, $0.3$ and $0.4$. The uniform magnetic 
flux $\phi_{\rm uni}$ is scaled by the strength of the random 
magnetic flux $\sqrt{\langle \phi^2 \rangle}$. Inset: The normalized 
conductivity $\sigma /\sigma_0$ plotted as a function of 
$\phi_{\rm uni}\cdot\sigma_0$. From the top, results for 
$h_{\rm rf}=0.4$, $0.3$ and $0.2$ are presented. 
The normalized data do not fall on the one curve, suggesting 
the deviation from the Drude behavior. \label{fig4}}
\end{figure}

In order to see whether this peak structure around zero field can be 
observed in the presence of the random potential $(W\neq 0)$, we 
also examine the electron diffusion with the additional random potential.
We consider the case where $h_{\rm rf}=0.2$ and $W=0$, $1$ and $2$ 
and find that the peak structure 
disappears at $W=2$ (Fig. 5). The random potential 
$W=2$ on the square lattice 
yields the mean free path on the order of 4 lattice 
constants.\cite{ESZ} This result indicates that to observe 
this enhancement of conductivity at zero uniform field
we need fairly clean samples. 

\begin{figure}
\includegraphics[scale=0.7]{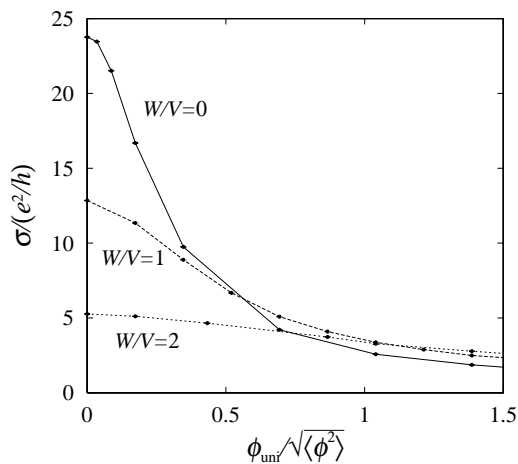}%
\caption{Conductivity in units of $e^2/h$ for 
$h_{\rm rf}=0.2$ and $W/V=0,1$ and 
$2$. \label{fig5}}
\end{figure}

Estimation of the mean free path in the random magnetic
field, especially for $W=0$, is a subtle problem.\cite{AMW} 
If we define the relaxation time for the transport\cite{AMW} by
$\tau_{\rm tr}=2D/v_{\rm F}^2$, where $v_{\rm F}$ denotes the Fermi velocity, 
the mean free path $l_{\rm rf}$ in 
the random magnetic field can be defined by $l_{\rm rf}= v_{\rm F} 
\tau_{\rm tr}
= 2D/v_{\rm F}$. Near the band center, the average Fermi velocity 
can be estimated to be on the order of 
$2aV/\hbar$. The mean free path $l_{\rm rf}$ 
is then evaluated to be about $21a$, $9a$, and $5a$ for
$h_{\rm rf}=0.2$, $0.3$ and $0.4$, respectively.

\section{Discussion}

We have demonstrated 
that in the absence of the uniform magnetic field, 
the conductivity is inversely proportional to the square of the magnitude 
of the random field, and hence is very sensitive to the strength  of the 
random magnetic field. 
In contrast, we have also shown that
the conductivity takes a value on the order of $e^2/h$ and is  
insensitive to the strength of the random magnetic field if the uniform 
field is set to be 
$\phi_{\rm uni}\approx \sqrt{\langle \phi^2 \rangle}$ (Fig. 4).
These two properties yield the peak structure at zero field, especially, 
in the case of the weak random magnetic fields.

It may be useful to 
discuss our results in view of the Drude formula $\sigma = \sigma_0 /
(1+\omega_c^2 \tau^2)$, where $\omega_c$ and $\tau$ are the 
cyclotron frequency and the relaxation time, respectively.
Our results indicate that 
$\sigma_0 \propto \tau \propto \langle\phi^2\rangle^{-1}$(Fig. 2).
The Drude theory then yields the form $\sigma /\sigma_0 =
1/(1+A(\phi_{\rm uni}\cdot \sigma_0)^2)$, where $A$ is a constant 
fixed by the density of electrons and 
independent of $\phi_{\rm uni}$ and $\phi$. We have found that 
although 
the data for small magnetic fields $ \phi_{\rm uni} \ll \sqrt{\langle \phi^2 
\rangle }$ seem to be 
consistent with this scaling form, it is unlikely that
the Drude formula accounts for the behavior around $\phi_{\rm uni} \approx 
\sqrt{\langle \phi^2 \rangle }$.
The deviation from the 
Drude theory, in which the resistivity is independent of the magnetic field,  
comes from the fact that the conductivity is 
insensitive to the magnitude of the random magnetic field at 
$\phi_{\rm uni}\approx \sqrt{\langle \phi^2 \rangle}$ (Fig.4).

Let us consider this insensitivity of 
the conductivity to the magnitude of the 
random magnetic field at 
$\phi_{\rm uni}\approx \sqrt{\langle \phi^2 \rangle}$ 
in more detail. With this condition 
$\phi_{\rm uni}\approx \sqrt{\langle \phi^2 \rangle}$, the system is 
almost equivalent to a system having
the random magnetic fields distributed in the range $0 \leq \phi / 
\phi_0 \leq h$ and having 
no uniform magnetic field $\phi_{\rm uni} = 0$.
We have then evaluated  the conductivity 
in such a system for 
$0.2$, $0.3$, $0.4$ and $0.5$ and, indeed, 
found that for all these values of $h$ the 
conductivity is insensitive to the value of $h$ and 
falls in the range $1.4 \sim 1.9 (e^2/h)$.
This would be one of the significant transport properties 
specific to the random magnetic fields.

In order to consider a possible relationship to the 
experiments, it may be useful to identify 
the correlation length of the random fields in 
experiments with the lattice constant of the present model.
Our calculation then implies that the sample having 
mean free paths longer than 4 correlation 
lengths of the random fields is very sensitive to the 
application of uniform magnetic fields. This implication seems to be 
consistent with the 
experiments\cite{Ando} where the dip structure is observed 
in samples having a mean free path larger than the correlation 
length of the random magnetic field.

In summary, we have investigated the electron diffusion 
in the random magnetic field with additional uniform 
magnetic fields by the equation of motion 
method. We have found that a sharp peak at zero uniform field
appears in magnetoconductance in the 
absence of the spatial correlation of 
random magnetic fields, where the semi-classical 
theory can not be applied. The width of the peak turns out to be 
on the order of the fluctuation of the random magnetic field. 
The conductivity at $\phi_{\rm uni}\approx 
\sqrt{\langle \phi^2 \rangle}$ is found to be insensitive to 
the magnitude of the random magnetic field.
This peak structure disappears when the mean free path 
becomes shorter by introducing the 
random potential. 
Although the present method enables us to simulate very large 
systems, we can obtain the longitudinal conductivity $\sigma_{xx}$ only.
Detailed analysis of $\sigma_{xx}$ in experiments is 
required to examine the relevance of the present results.

\begin{acknowledgments}
The authors thank M. Ando, Y.Ono, B.Kramer and  S. Kettemann for valuable 
discussions. Numerical calculations were performed by the facilities 
of the Supercomputer Center, Institute for Solid State Physics, 
University of Tokyo. This work is partly supported by the 
Grants-in-aid No.13740244 from Japan Society for the Promotion of Science.
\end{acknowledgments}


\begin{thebibliography}{10}
%
\bibitem{HLR}
B. I. Halperin, P. A. Lee, and N. Read, Phys. Rev. B {\bf 47},  7312  (1993).

\bibitem{LR}
P. A. Lee $\&$ T. V. Ramakrishnan, Revs. Mod. Phys. {\bf 57},  287  (1985).

\bibitem{PZ}
C. Pryor and A Zee, Phys. Rev. B {\bf 46}, 3116 (1992).
\bibitem{KZ}
V. Kalmeyer and S.C. Zhang, Phys. Rev. B {\bf 46}, 9889 (1992).
\bibitem{SN}
T. Sugiyama and N. Nagaosa, Phys. Rev. Lett. {\bf 70}, 1980 (1993).
\bibitem{AHK}
Y. Avishai, Y. Hatsugai and M. Kohmoto, Phys. Rev. B {\bf 47}, 9561 (1993).
\bibitem{YG}
K. Yakubo and Y. Goto, Phys. Rev. B {\bf 54}, 13432 (1996).
\bibitem{Verges}
J.A. Verg\'{e}s, Phys. Rev. B {\bf 57}, 870 (1998).
\bibitem{SYN}
H. Shima, K. Yakubo and T. Nakayama, J. Phys. Soc. Jpn. 
{\bf 70}, 2682 (2001).
\bibitem{Yakubo}
K. Yakubo, J. Phys. Soc. Jpn. {\bf 70}, 3331 (2001).
\bibitem{KO} T. Kawarabayashi and T. Ohtsuki, Phys. Rev. B{\bf 51},
10897 (1995).

\bibitem{OSO} T. Ohtsuki, K. Slevin and Y. Ono, J. Phys. Soc. Jpn. 
{\bf 62}, 3979 (1993).

\bibitem{Furusaki} 
A. Furusaki, Phys. Rev. Lett. {\bf 82},  604  (1999).

\bibitem{MBF} 
C. Mudry and P. W. Brouwer, and A. Furusaki, Phys. Rev. B {\bf 59},  13221  
(1999).
\bibitem{EMPW}
F. Evers, A. D. Mirlin, D. G. Polyakov, and P. W\"{o}lfle, Phys. Rev. 
B{\bf 60},  8951  (1999).

\bibitem{Ando}
M. Ando, A. Endo, S. Katsumoto, and Y. Iye, Physica B{\bf 284}-{\bf 288},
1900 (2000); M. Ando, Thesis (2001).

\bibitem{MWEPW}
A.D. Mirlin, J. Wilke, F. Evers, D.G. Polyakov and P. W\"{o}lfle, 
Phys. Rev. Lett. {\bf 83}, 2801 (1999); F. Evers, A.D. Mirlin, 
D.G. Polyakov and P. W\"{o}lfle, Phys. Rev. B{\bf 60}, 8951 (1999);
A.D. Mirlin, D.G. Polyakov and P. W\"{o}lfle, Phys. Rev. Lett. {\bf 80},
2429 (1998).





\bibitem{Suzuki}
M. Suzuki, Phys. Lett. A {\bf 146}, 319 (1990). 

\bibitem{KO2} T. Kawarabayashi and T. Ohtsuki, Phys. Rev. B{\bf 53},
6975 (1996).
\bibitem{KTH} R. Kubo, M. Toda, and N. Hashitsume, {\it Statistical 
Physics II}, 2nd ed. (Springer-Verlag, Berlin, 1991).
\bibitem{SKM} L. Schweitzer, B. Kramer and A. MacKinnon, J. Phys. C{\bf 17},
4111 (1984).

\bibitem{ESZ} E.N. Economou, C.M. Soukoulis and A.D. Zdetsis, 
              Phys. Rev. B{\bf 30}, 1686 (1984).

\bibitem{AMW} A.G. Aronov, A.D. Mirlin and P. W\"{o}lfle, Phys.
Rev. B{\bf 49}, 16609 (1994).

\end{thebibliography}
\end{document}